\documentclass[twocolumn,aps,pra,preprintnumbers,bibnotes10pt,superscriptaddress,amsmath,amssymb]{revtex4-2}
\usepackage{graphicx,subfigure,wrapfig}
\usepackage{float}
\usepackage{color}
\usepackage{epstopdf}
\usepackage{amsmath}
\usepackage{braket}
\usepackage{amsthm}
\usepackage{amssymb}
\usepackage{amsfonts}
\usepackage{graphicx,subfigure}
\usepackage{txfonts}
\usepackage{ulem}
\usepackage{mathtools}
\usepackage{bm}
\usepackage{hyperref}
\usepackage{lineno}
\usepackage{natbib}
\usepackage{comment}

\newcommand{\mean}[1]{\ensuremath{\big\langle #1 \big\rangle}}

\newcommand{\be}{\begin{equation}}
\newcommand{\ee}{\end{equation}}
\newcommand{\beq}{\begin{eqnarray}}
\newcommand{\eeq}{\end{eqnarray}}
\newcommand{\vect}[1]{\bm{#1}}

\newcommand{\thetat}{\theta_{\rm true}}
\newcommand{\sigmat}{\sigma_{\rm true}}
\newcommand{\taut}{\tau_{\rm true}}
\newcommand{\thetae}{\theta_{\rm est}}
\newcommand{\sigmae}{\sigma_{\rm est}}
\newcommand{\taue}{\tau_{\rm est}}
\newcommand{\Xit}{\vect{\Xi}_{\rm true}}
\newcommand{\Xie}{\vect{\Xi}_{\rm est}}

\begin{document}


\title{Joint estimation of phase and uncorrelated dephasing \\ 
in a differential quantum interferometer} 

\author{L. Pezzè}
\email{luca.pezze@ino.it}
\affiliation{Istituto Nazionale di Ottica, Consiglio Nazionale delle Ricerche (INO-CNR), Largo Enrico Fermi 6, 50125 Firenze, Italy}
\affiliation{European Laboratory for Nonlinear Spectroscopy (LENS), Via N. Carrara 1, 50019 Sesto Fiorentino, Italy}
\affiliation{QSTAR, Largo Enrico Fermi 2, 50125 Firenze, Italy}

\author{A. Santoni}
\affiliation{European Laboratory for Nonlinear Spectroscopy (LENS), Via N. Carrara 1, 50019 Sesto Fiorentino, Italy}
\affiliation{University of Naples “Federico II”, Via Cinthia 21, 80126 Napoli, Italy} 

\author{C. Mazzinghi} 
\affiliation{Istituto Nazionale di Ottica, Consiglio Nazionale delle Ricerche (INO-CNR), Largo Enrico Fermi 6, 50125 Firenze, Italy}
\affiliation{European Laboratory for Nonlinear Spectroscopy (LENS), Via N. Carrara 1, 50019 Sesto Fiorentino, Italy}

\author{M. Fattori}
\affiliation{Istituto Nazionale di Ottica, Consiglio Nazionale delle Ricerche (INO-CNR), Largo Enrico Fermi 6, 50125 Firenze, Italy}
\affiliation{European Laboratory for Nonlinear Spectroscopy (LENS), Via N. Carrara 1, 50019 Sesto Fiorentino, Italy}
\affiliation{University of Florence, Physics Department, Via Sansone 1, 50019 Sesto Fiorentino, Italy}

\author{A. Smerzi}
\affiliation{Istituto Nazionale di Ottica, Consiglio Nazionale delle Ricerche (INO-CNR), Largo Enrico Fermi 6, 50125 Firenze, Italy}
\affiliation{European Laboratory for Nonlinear Spectroscopy (LENS), Via N. Carrara 1, 50019 Sesto Fiorentino, Italy}
\affiliation{QSTAR, Largo Enrico Fermi 2, 50125 Firenze, Italy}

\begin{abstract}

Precise measurements in optical and atomic systems often rely on differential interferometry. 
This method allows to handle large and correlated phase noise contributions -- such as environmental vibrations, thermal fluctuations, or instrumental drifts -- preventing them from blurring the signal.
To date, this approach has primarily focused on extracting the differential phase shift. 
However, valuable information about the system is also contained in the width of uncorrelated phase fluctuations.
In this work, we present a maximum likelihood approach for the simultaneous estimation of both the differential phase shift and the width of uncorrelated phase noise. 
Unlike conventional methods, our technique explicitly accounts for the data spreading and outperforms traditional ellipse fitting in terms of both precision and accuracy. 
We demonstrate our methodology using a quantum mechanical model of coupled interferometers, where uncorrelated dephasing arises from projection noise and interparticle interactions.
Our results establish a novel approach to data analysis in differential interferometry that is readily applicable to current experiments.

\end{abstract}

\maketitle 

\section{Introduction} 

Differential interferometry is a powerful and widespread technique in atomic sensing~\cite{BongsNRV2019,GeigerAVS2020,NarducciAPX2022}.
In this approach, two or more interferometers operate in parallel, in a configuration that guarantees a common-mode phase noise.
By leveraging the correlations between the output measurements, the differential phase shift can be extracted regardless the noise strength.
This approach is exploited in a broad range of applications in inertial sensors~\cite{SnaddenPRL1998, GauguetPRA2009, SorrentinoPRA2014, BarrettAVS2022,GersemanEPJD2020,JanvierPRA2022} and atomic clocks~\cite{YoungNATURE2020, ZhengNATURE2022, EcknerNATURE2023}, including tests of the equivalence principle~\cite{ZhouPRL2015, AsenbaumPRL2020}, measurement of physical constants~\cite{FixlerSCIENCE2007, LamporesiPRL2008, RosiNATURE2014, ParkerSCIENCE2018}, and gravity mappings~\cite{StrayNATURE2022}. 

A key aspect of differential interferometry is the nontrivial parameter estimation analysis~\cite{FosterOPTLETT2002, StocktonPRA2007, PereiraPRA15, ZhangOPTEXP2023,RidleyEPJ2024}.  
When the mean signal from each interferometer varies sinusoidally with the phase shift -- as encountered in common setups -- the combined output from two noise-correlated interferometers form, in average, an ellipse~\cite{FosterOPTLETT2002}.
The shape and orientation of such ellipse depend on the differential phase shift.
As a result, ellipse fitting has become the most common approach to phase estimation in differential interferometry. 
However, even with perfect common-mode noise correlations between the two sensors, and in the absence of any technical imperfections (such as fluctuations of offset and contrast of the interference fringes), the intrinsic quantum noise in each interferometer acts as uncorrelated dephasing.
This leads to an unavoidable spreading of measurement data around the average ellipse. 
This diffusion impacts the uncertainty in estimating the differential phase signal using ellipse-fitting approaches, and may introduce estimation biases that are generally difficult to quantify and correct.
Furthermore, conventional data analyses typically extract only the differential phase shift while overlooking valuable information about noise and imperfections contained in the measurement distribution.
In addition to the projection noise mentioned above, other noise sources may arise from particle-particle interactions and classical dephasing (e.g. due to Raman laser frequency noise or magnetic field disturbances), which can be useful to estimate -- for instance, to characterize the operations of the interferometric system.

In this manuscript, we propose a multiparameter maximum likelihood analysis to estimate -- simultaneously, from the same set of correlated data -- the differential phase signal, $\theta$, and the width, $\sigma$, of the uncorrelated  dephasing.
The multiparameter approach intrinsically accounts for the spreading of measurement data and provides improved precision and accuracy in the estimation of $\theta$, compared to ellipse fitting.
In particular, the ML approach saturates the multiparameter Cram\'er-Rao bound, which we compute analytically in suitable limits:   
\be \label{results}
\Delta \theta = \frac{\sigma}{\sqrt{m}}, \quad {\rm and} \quad \Delta \sigma = \frac{\sigma}{\sqrt{2m}},
\ee
where $m$ is the number of joint measurements in the two interferometers. 
In particular, Eq.~(\ref{results}) clearly shows that the uncorrelated noise determines the accuracy with which $\theta$ is estimated.

We apply our method to a differential sensor made of two quantum Mach-Zehnder interferometers~\cite{EckertPRA2006, LandiniNJP2014, CorgierQUANTUM2023, CorgierARXIV} using interacting particles. 
In this case, 
\be \label{sigmaintro}
\sigma^2 = \frac{2}{N} + 2 N \tau^2,
\ee
where the first term is due to projection noise, with $N$ being the number of particles in each interferometer, and the second term is due to interparticle interactions, with $\tau = gT/\hbar$, $g$ being the interaction strength and $T$ the interrogation time.
Our approach enables the joint estimation of $\tau$ (or, equivalently, $g$) and $\theta$. 
Estimating the strength of particle-particle interaction can be useful to locale, with high precision, the zero crossing of a Feshbach resonance~\cite{FattoriPRL2008, SpagnolliPRL2018}. 
Tuning the interatomic interactions then allows to increase the sensitivity of the device by extending its coherence time~\cite{Petrucciani}. 

The joint estimation of signal and noise in a quantum sensor has garnered significant attention in the literature~\cite{VidrighinNATCOMM2014, CrowleyPRA2014, BelliardoNJP2024, SzczykulskaQST2017, AltorioPRA2015, BirchallPRL2020}, in the context of multiparameter estimation~\cite{PezzeARXIV}.
Here we address this problem in the  experimentally-relevant framework of differential interferometry.

\section{Quantum description of a differential interferometer}

\begin{figure}[t!]
\includegraphics[width=0.9\columnwidth]{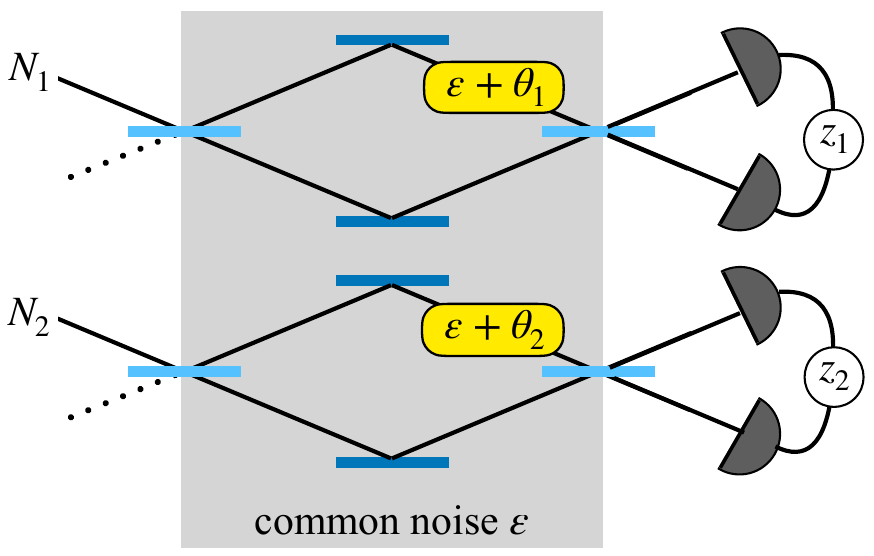}
\caption{Scheme of the differential interferometer studied in this manuscript.
It consists of two Mach-Zehnder interferometers operating in parallel and affected by a common phase noise $\varepsilon$ with shot-to-shot fluctuations. 
Interacting particles enter a single input port of each interferometer, while the phase is estimated from the measurement of the normalized relative number of particles, $z_1$ and $z_2$, in each interferometer.
}
\label{Figure0}
\end{figure}

Our differential sensing scheme consists of two Mach-Zehnder (or Ramsey) interferometers working in parallel and affected by a common phase noise, see Fig.~\ref{Figure0}.
In the following, we first recall the theoretical description of a single quantum interferometer, e.g. following Refs.~\cite{YurkePRA1986, PezzeRMP2018}, and then discuss the differential scheme.

\subsection{Single quantum interferometer}

We introduce effective collective spin operators $\hat{J}_{x} = (\hat{a}^\dag \hat{b} + \hat{b}^\dag \hat{a})/2$, $\hat{J}_{y} = (\hat{a}^\dag \hat{b} - \hat{b}^\dag \hat{a})/(2i)$ and $\hat{J}_{z} = (\hat{a}^\dag \hat{a} - \hat{b}^\dag \hat{b})/2$, where $\hat{a}$ and $\hat{b}$ ($\hat{a}^\dag$ and $\hat{b}^\dag$) are bosonic annihilation (creation) operators for the modes $a$ and $b$.
The interferometric sequence starts with all $N$ particles in mode $a$: we indicate this spin-polarized state as $\ket{\psi_0}$.
Each interferometer is realized by a sequence of three transformations: a beam splitter, a phase acquisition stage, of duration time $T$, and a final beam splitter.
The first balanced beam splitter is describes by $e^{-i (\pi/2)\hat{J}_{y}}$.
For $N\gg 1$, the quantum state at this stage can be approximated by a Gaussian distribution
\be \label{psiBS}
\ket{\psi_{{\rm BS}}} = e^{-i (\pi/2)\hat{J}_{y}} \ket{\psi_0} = \sum_{\mu=-N/2}^{N/2} \frac{e^{-\mu^2/N}}{(\pi N/2)^{1/4}} \ket{\mu},
\ee
where $\ket{\mu}$ is the eigenstate of $\hat{J}_{z}$ with eigenvalue $\mu=-N/2, -N/2+1, ..., N/2$.
After the beam splitter, during phase encoding, modeled as $e^{-i \varphi \hat{J}_{z}}$, we turn on the particle-particle interaction, described by $e^{-i \tau \hat{J}_{z}^2}$~\cite{notaBS}.
Using Eq.~(\ref{psiBS}), the quantum state after phase encoding is thus given by
\beq
\ket{\psi_{{\rm PS}}} = \sum_{\mu=-N/2}^{N/2} \frac{e^{-\mu^2/N}}{(\pi N/2)^{1/4}} e^{-i \varphi \mu} e^{-i \tau \mu^2} \ket{\mu},
\eeq
We can evaluate the uncorrelated noise due to intrinsic quantum fluctuations (projection noise) and interaction by following Ref.~\cite{JavanainenPRL1994} and projecting $\ket{\psi_{{\rm PS}}}$ over phase states
\be
\ket{\phi} = \frac{1}{\sqrt{N+1}} \sum_{\mu=-N/2}^{N/2} e^{-i \phi \mu} \ket{\mu}.
\ee
By replacing the sum with an integral (valid for $N\gg 1$), we compute the phase distribution 
\be \label{Pphasedist}
P(\phi) = \vert \bra{\phi} \psi_{{\rm PS}} \rangle \vert^2 = \sqrt{\frac{1}{2 \pi \sigma^2}} e^{-\tfrac{(\phi-\varphi)^2}{2 \sigma^2_\phi}},
\ee
with dephasing rate~\cite{JavanainenPRL1994} 
\be \label{dephasingint}
\sigma^2_{\phi} = \frac{1}{N} + N \tau^2.
\ee
Equation (\ref{dephasingint}) is valid when $\sigma^2_\phi \ll \pi$.
Alternatively, we can estimate the width of the phase noise by using a mean-field (or Holstein-Primakof~\cite{PezzeRMP2018}) approximation obtained by replacing the mode operators with complex numbers $\hat{a} \sim \sqrt{N_a}e^{i\varphi_a}$ and $\hat{b} \sim \sqrt{N_b}e^{i\varphi_b}$, where $N_{a}=\mean{\hat{a}^\dag \hat{a}}$ and $N_{b}=\mean{\hat{b}^\dag \hat{b}}$.
We thus obtain $\hat{J}_y \sim \sqrt{N_a N_b} \sin(\varphi_a-\varphi_b)$. 
For the state $\ket{\psi_{{\rm PS}}}$, we have $N_a = N_b = N/2$, and we can identify $\phi_a-\phi_b$ as the dephasing between the two modes.
Assuming $\phi_a-\phi_b \ll \pi$ [such that $\sin (\phi_a-\phi_b) \approx \phi_a-\phi_b$], we identify the dephasing rate as
\be \label{sigmaphiapp}
\sigma^2_{\phi} = \frac{4}{N^2} (\Delta \hat{J}_y)^2, 
\ee
where~\cite{KitagawaPRA1993}
\be \label{DJyKU}
(\Delta \hat{J}_y)^2  
= \frac{N}{4}\Big(1+ \frac{N-1}{4}\big[A + \sqrt{A^2 + B^2} \cos(2\delta) \big] \Big),
\ee
$A = 1 - \cos^{N_j-2}(2\tau)$, $B=4 \sin(\tau)\cos^{N-2}(\tau)$ and $\delta = \tfrac{1}{2} \arctan \tfrac{B}{A}$.
A Taylor expansion of Eqs.~(\ref{sigmaphiapp}) and (\ref{DJyKU}) for $\tau \to 0$ gives $\sigma^2_{\phi} \approx 1/N + [(N-1)^2/N_j] \tau^2$, which agrees with Eq.~(\ref{dephasingint}) for $N \gg 1$ and small $\tau$.
The final beam splitter of each interferometer is described by $e^{-i (\pi/2)\hat{J}_x}$~\cite{footnoteBS}.
We measure the relative number of particles $\hat{J}_z$ on the output state $\ket{\psi_{{\rm MZ}}} = e^{-i (\pi/2)\hat{J}_x} \ket{\psi_{{\rm PS}}}$, with possible result $\mu \in [-N/2, N/2]$.
We indicate with $z = 2 \mu/N \in [-1, 1]$ the normalized relative number of particles.
Its mean value $\bar{z} = 2\langle \hat{J}_{z} \rangle/N$ is
\be \label{zave}
\bar{z} = V(\tau) \sin(\varphi), 
\ee
where $V(\tau) = 2\bra{\psi_0} \hat{J}_{x} \ket{\psi_0}/N = \cos^{N-1}(\tau) \approx 1 -  \tfrac{N-1}{2}\tau^2$ is the visibility of the interference signal.
We assume $\tau^2 N \ll 1$ and thus neglect the loss of visibility due to interaction, namely $V(\tau) \approx 1$.

\subsection{Differential scheme}

In the differential scheme, see Fig.~\ref{Figure0}, the single-shot phase acquired in each interferometer (labeled by $j=1,2$) is $\varphi_1 = \varepsilon + \theta_1$ and $\varphi_2 = \varepsilon + \theta_2$, where $\varepsilon$ is a fluctuating phase due to noise, while $\theta_1$ and $\theta_2$ are constant phase shifts.
We consider $\varepsilon$ with shot-to-shot variations with a uniform distribution $\mathcal{P}_{\rm cn}(\varepsilon)=1/(2\pi)$ spanning the full $[-\pi, \pi]$ interval. 
In atom interferometers, correlated phase noise can be associated to the laser source phase, vibration noise, and light shift noise.
%
According to the Eq.~(\ref{zave}), we have that the average number of particles at the output of each interferometer is  
\begin{subequations} \label{Eqz}
\begin{align}
& \bar{z}_1 = \sin (\varepsilon + \theta_1), \label{Eqz1} \\
& \bar{z}_2 = \sin (\varepsilon + \theta_2). \label{Eqz2}
\end{align}
\end{subequations}
Due to correlations between measurement data established by $\varepsilon$, the average values in Eq.~(\ref{Eqz}) distribute along the ellipse 
\be \label{ellipse}
\bar{z}_1^2 + \bar{z}_2^2 -2 \bar{z}_1 \bar{z}_2 \cos \theta  - \sin^2 \theta = 0,
\ee
where 
$\theta = \theta_1 - \theta_2$ is the differential phase shift.
Yet, single measurements results $z_1$ and $z_2$ spread out from the average ellipse according to the distribution
\be \label{Pz1z2quantum}
P(z_1,z_2\vert \theta, \sigma) = \int_{-\pi}^{\pi} \frac{d\epsilon}{2\pi}~P_1(z_1\vert \epsilon + \theta_1, \sigma_{\phi,1}) P_2(z_2\vert \epsilon + \theta_2, \sigma_{\phi,2}).
\ee 
Here, $P_1(z_1\vert \epsilon +\theta_1,\sigma_{\phi,1}) = \vert \langle z_1 \ket{\psi_{{\rm MZ}}(\epsilon +\theta_1, \sigma_{\phi,1})} \vert^2$, $P_2(z_2\vert \epsilon +\theta_2, \sigma_{\phi,2}) =\vert \bra{z_2} \psi_{2}(\epsilon +\theta_2, \sigma_{\phi,2})\rangle \vert^2$ are probabilities to obtain measurement result $z_j = 2\mu_j/N_j$ (j=1,2), where $\mu_j$ and $N_j$ are relative and total number of particles in the $j$th interferometer, and $\ket{z_j} \equiv \ket{\mu_j}$.
Notice that, due to the noise term $\epsilon$ being uniformly distributed in $[-\pi, \pi]$, Eq.~(\ref{Pz1z2quantum}) only depends on $\theta$ and the sum of uncorrelated noise terms, 
$\sigma^2 = \sigma_{\phi,1}^2+\sigma_{\phi,2}^2$.  
We recover Eq.~(\ref{sigmaintro}) by
using Eq.~(\ref{dephasingint}) and taking the number of particles and $\tau$ to be the same in both interferometers.

In principle, it would be possible to perform a maximum likelihood or a Bayesian joint estimation of $\theta$ and $\sigma$ by using the probability distribution Eq.~(\ref{Pz1z2quantum}).
However, experimentally, $P(z_1,z_2\vert \theta, \sigma)$ is accessed by a calibration of the differential interferometer that requires to acquire the statistics of $z_1$ and $z_2$ results for fixed, known, values of $\theta$ and $\sigma$. 
In practice, this task is very challenging because of the large number of particles and the multiple parameters.
Any parameter estimation approach that uses the full knowledge of Eq.~(\ref{Pz1z2quantum}) is thus unfeasible, in practice.

The most common approach to data analysis in differential interferometers consists of extracting $\theta$ by finding the ellipse line that better interpolates the sequence of $m$ measured data $\vect{z}_1=\{z_1^{(1)},..,z_1^{(m)}\}$ and $\vect{z}_2=\{z_2^{(1)},..,z_2^{(m)}\}$~\cite{FosterOPTLETT2002}.
Ellipse fitting has two limitations.
First, it only allows to estimate $\theta$.
Second, it is not guaranteed that ellipse fitting provides an optimal estimate of $\theta$, namely with the smallest possible uncertainty given the measurement observable.
Furthermore, the estimate can be biased due to the nonlinear fitting process and the spreading of data around the average ellipse Eq.~(\ref{ellipse})~\cite{CorgierARXIV}. 

In contrast, following Eq.~(\ref{Pphasedist}), our idea is to model the intrinsic quantum noise of the differential scheme as a classical uncorrelated noise with a Gaussian distribution of width $\sigma^2$.
This provides an approximation of Eq.~(\ref{Pz1z2quantum}) that is suitable to extract both $\theta$ and $\sigma$ simultaneously, using a maximum likelihood approach, see Sec.~\ref{Sec3}.
We show that our approach is characterized by a better precision than ellipse fitting and has negligible bias, at least for an experimentally-relevant number of measurement data, see Sec.~\ref{Sec4}.  

\begin{figure*}[t!]
\includegraphics[width=0.68\textwidth]{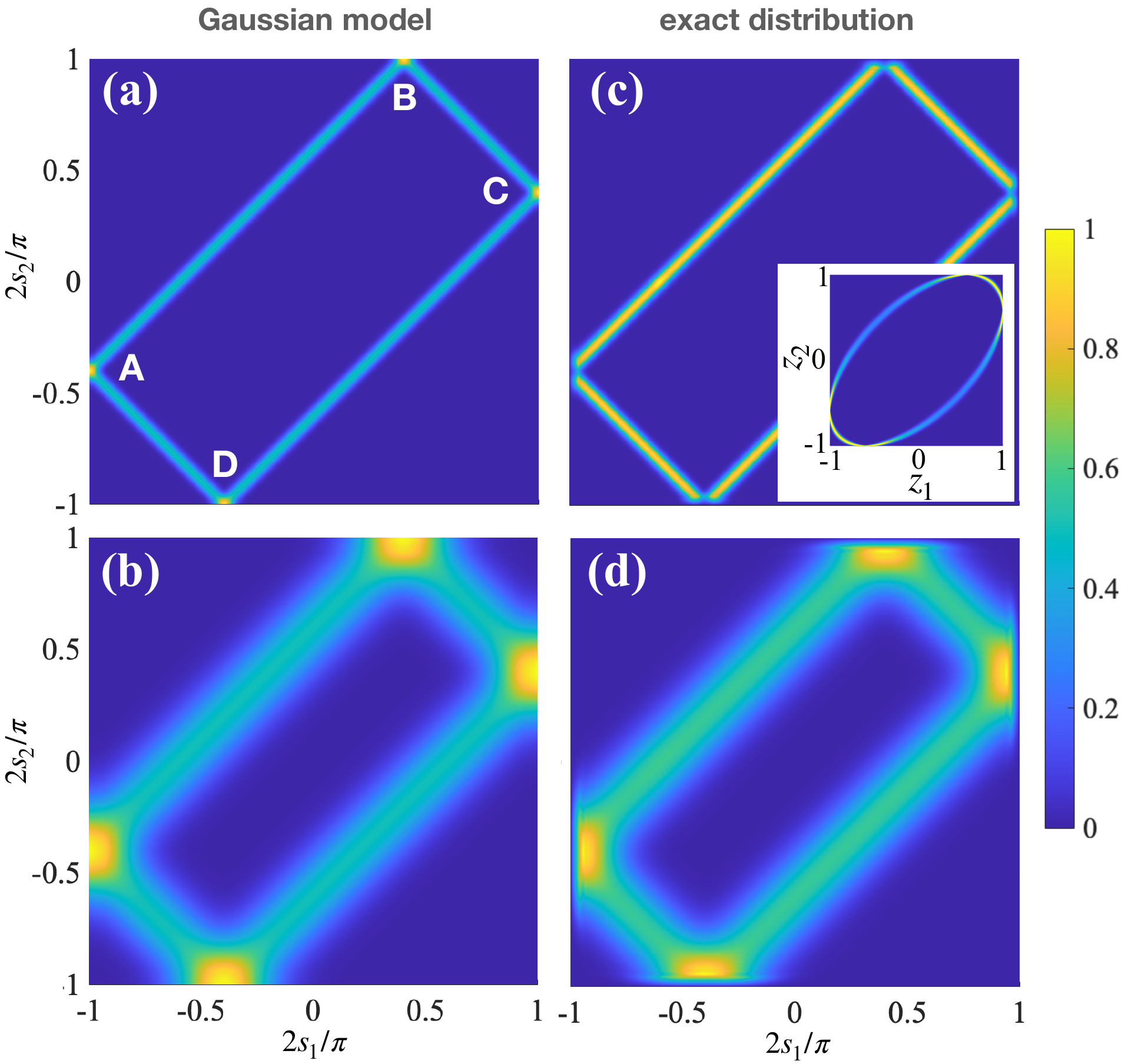}
\caption{
Probability density distribution $P_s(s_1, s_2 \vert \theta, \sigma)$ Eq.~(\ref{Eq.probability2}) obtained from the Gaussian model presented in Sec.~\ref{GNM}, and plotted as a function of $s_1$ and $s_2$ for (a) $\tau=0$ and (b) $\tau=0.005$.
In panel (a) we report the position of the edged A-D: Eq.~(\ref{Eq.cooA})-(\ref{Eq.cooD}).
For comparison, in panel (c) and (d), we report the corresponding exact distribution $P(s_1,s_2\vert \theta,\sigma)$ that can be obtained from Eq.~(\ref{Pz1z2quantum}) after a change of variables.
In the inset of panel (c) we show  Eq.~(\ref{Pz1z2quantum}) as a function of $z_1$ and $z_2$, showing the characteristic elliptic shape.
In all panels, $\theta/\pi=0.3$, $N=1000$, and the color scale 
corresponds to the probability density distribution normalized to its maximum value.
}
\label{Figure1}
\end{figure*}

\section{Classical gaussian model and MULTIPARAMETER MAXIMUM LIKELIHOOD aproach}
\label{Sec3}

\subsection{Classical Gaussian noise model}
\label{GNM}

We model stochastic shot-to-shot measurement results as  
\begin{subequations} \label{Eqzave}
\begin{align}
& z_1 = \sin (\varepsilon + \chi + \theta), \label{Eqz1} \\
& z_2 = \sin (\varepsilon),  \label{Eqz2}
\end{align}
\end{subequations}
where $\varepsilon$ is a common (correlated) phase noise uniformly distributed in $[-\pi,\pi]$, as considered above, and $\chi$ is an uncorrelated noise with Gaussian distribution 
\be \label{Pnoisedistuncorr}
\mathcal{P}(\chi) = \mathcal{N}e^{-\frac{\chi^2}{2\sigma^2}}.
\ee
In Eq.~(\ref{Pnoisedistuncorr}), $\mathcal{N}$ is the normalization and $\sigma$ is the uncorrelated noise width.
According to Eqs.~(\ref{Eqzave}) and (\ref{Pnoisedistuncorr}), $z_1$ and $z_2$ are treated as classical variables with fluctuations due to both correlated and uncorrelated phase noise. 

In the following, we characterize the conditional probability density $P_z\big(z_1, z_2 \vert \theta,\sigma \big)$ of $z_1$ and $z_2$ values. 
For convenience, we consider the variables $s_1 = {\rm arcsin}(z_1)$ and $s_2 = {\rm arcsin}(z_2)$, instead of $z_1$ and $z_2$, and directly compute $P_s\big(s_1, s_2 \vert \theta, \sigma \big)$.
The two distributions are related by the change of variables 
\be \label{Pz}
P_s(s_1,s_2\vert \theta,\sigma) = 
P_z\big(z_1, z_2 \vert \theta,\sigma \big) \sqrt{1-z_1^2} \sqrt{1-z_2^2}.
\ee
The inversion of the sinusoidal functions in Eq.~(\ref{Eqzave}) has a $\pi$ ambiguity. 
More explicitly, inverting Eq.~(\ref{Eqz1}) gives $\varepsilon +\chi + \theta = s_1$ and $\varepsilon +\chi + \theta = \pi - s_1$, while inverting Eq.~(\ref{Eqz2}) gives $\varepsilon = s_2$ and $\varepsilon = \pi - s_2$.
These four equivalent possibilities are weighted according to $\mathcal{P}(\chi)$.
The probability density $P_s\big(s_1, s_2 \vert \theta,\sigma \big)$ thus consists of four terms~\cite{notaP}:
\beq \label{Eq.probability2}
P_s\big(s_1, s_2 \vert \theta,\sigma \big)
&=& \frac{1}{2\pi} \Big(
\mathcal{P}\big(\theta - s_1 + s_2 \big) 
+  \mathcal{P}\big(\theta + \pi - s_1 - s_2 \big) + \nonumber \\
&& 
+  \mathcal{P}\big(\theta - \pi + s_1 + s_2 \big) +  \mathcal{P}\big(\theta + s_1 - s_2 \big) \Big).
\eeq
Equation~(\ref{Eq.probability2}) has a characteristic rectangular shape, see Fig.~\ref{Figure1}(a) and (b). 
This can be understood by taking $\sigma=0$, which corresponds to the delta function $\mathcal{P}(\chi) = \delta(\chi)$.
In this case, Eq.~(\ref{Eq.probability2}) predicts that the possible values of $s_1$ and $s_2$ follow the four lines $s_1 - s_2 = \pm \theta$ and $s_1 + s_2 = \pm (\theta-\pi)$.
The edges $\{s_1,s_2\}$ of the rectangle, where two lines cross, are 
\begin{subequations} \label{Eq.coo}
\begin{align}
A &= \{s_{1}=-\pi/2+\theta, s_2=-\pi/2\}, \label{Eq.cooA} \\
B &= \{-\pi/2, -\pi/2 + \theta\}, \label{Eq.cooB} \\
C &= \{\pi/2-\theta, \pi/2\}, \label{Eq.cooC} \\
D &= \{ \pi/2, \pi/2-\theta \}, \label{Eq.cooD} 
\end{align}
\end{subequations}
which are fully determined by $\theta$.
For the special case $\theta=0$ ($\theta=\pi$), we have $A=B$ and $C=D$ ($A=D$ and $B=C$): the rectangle degenerates into the line $s_1-s_2=0$ ($s_1 + s_2 = 0$).
In the case $\sigma_{\rm uncorr} \neq 0$, we finds that the Eq.~(\ref{Eq.probability2}) is a spread rectangle, see Fig.~\ref{Figure1}(a).
At the edges, the probability of $s_2$ (or $s_1$) values is approximately twice as large as the probability along the arm of the rectangle due to the joint pairwise contribution of the different terms in Eq.~(\ref{Eq.probability2}).

\begin{figure*}[t!]
\includegraphics[width=1\textwidth]{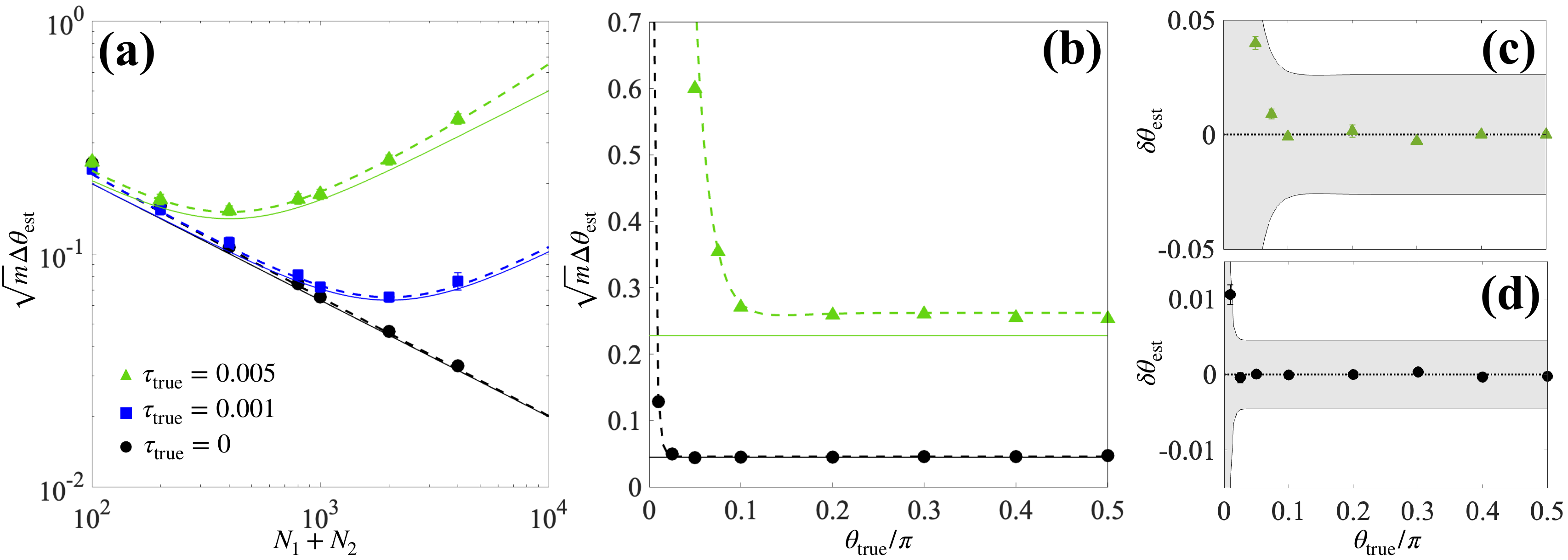}
\caption{(a) Phase estimation uncertainty $\Delta \thetae$ as a function of $N_1+N_2$.
Different symbols refer to different values of $\tau_{\rm true}$ (see legend).
For each value of $\tau_{\rm true}$, the dashed line is the CRB, while the solid line is Eq.~(\ref{Deltatheta}).
Here, $\thetat/\pi = 0.5$.
(b) Uncertainty $\Delta \thetae$ as a function of $\thetat$ for $\taut = 0.005$ (triangles) and $\taut=0$ (dots).
The dashed line the CRB, while the solid line is Eq.~(\ref{Deltatheta}).
Panels (c) and (d) show the bias $\delta\thetae$ as a function of $\thetat$.
The gray area corresponds to a region of width equal to the CRB around $\delta\thetae=0$ (dotted line).  
In all panels $N_1=N_2 = 1000$ and $m=100$.
}
\label{Figure6}
\end{figure*}

In Fig.~\ref{Figure1} we show that Eq.~(\ref{Eq.probability2}) agree very well with the probability $P(s_1,s_2\vert \theta,\sigma)$ obtained from Eq.~(\ref{Pz1z2quantum}) after a change of variables. 
In the figure, both distributions are calculated for the same $\theta$ and $\sigma$.
The agreement is due to the Gaussian nature of the uncorrelated noise, which can be well reproduced by the model in Eqs.~(\ref{Eqzave}) and (\ref{Pnoisedistuncorr}).
The two distributions $P(s_1,s_2\vert \theta,\sigma)$ and $P_s(s_1,s_2\vert \theta,\sigma)$ mainly differ close to the edges $2s_{1,2}/\pi = \pm 1$. 
For clarity, in the inset of Fig.~\ref{Figure1}(c) we plot $P(z_1,z_2\vert \theta,\sigma)$, Eq.~(\ref{Pz1z2quantum}), showing the characteristic elliptic distribution discussed above.

\subsection{Multiparameter Maximum Likelihood analysis}
\label{Sec.Multiparameter}

For clarity sake, indicate as $\Xit = \{\thetat, \sigmat\}$ the actual true value of the parameters we want to estimate, and with $\Xie = \{\thetae,\sigmae\}$ the estimated quantities. 
Within our maximum likelihood (ML) approach, 
\be \label{likelihood}
\Xie = {\rm argmax}_{\vect{\Xi}} \prod_{j=1}^m P_s(s_1^{(j)}, s_2^{(j)} \vert \vect{\Xi}), 
\ee 
where $\vect{\Xi} = \{\theta, \sigma\}$, $\vect{s}_1 = {\rm arcsin}(\vect{z}_1)$ and $\vect{s}_2 = {\rm arcsin}(\vect{z}_2)$ are the $m$ measured data for each interferometer.
The accuracy of the estimation is quantified by the $2\times 2$ covariance matrix 
\be \label{varest}
\vect{C} = \int d^m\vect{s}_1 d^m\vect{s}_2 ~ P_s(\vect{s}_1, \vect{s}_2 \vert \Xit) \big[\Xie(\vect{s}_1,\vect{s}_2)-\bar{\vect{\Xi}}_{\rm est}\big]^2,
\ee 
where 
\be \label{varmean}
\bar{\vect{\Xi}}_{\rm est} = \int d^m\vect{s}_1 d^m\vect{s}_2 ~  P_s(\vect{s}_1, \vect{s}_2 \vert \Xit) \Xie(\vect{s}_1,\vect{s}_2)
\ee
is the statistical mean value.
The diagonal elements of $\vect{C}$ give the variance for the estimation of the single parameters: specifically, $(\Delta \thetae)^2 = \vect{C}_{1,1}$ and $(\Delta \sigmae)^2 = \vect{C}_{2,2}$.
The off-diagonal element $\vect{C}_{1,2}$ gives statistical correlations between the $\thetae$ and $\sigmae$.
The precision of the estimation, given by the difference between the average estimate and the true value of the parameter,
\be \label{biasest}
\delta \vect{\Xi}_{\rm est} = \bar{\vect{\Xi}}_{\rm est} - \Xit. 
\ee

\subsection{Multiparameter Cram\'er-Rao bound}

The multiparameter Cram\'er-Rao bound (CRB) corresponding to the probability density distribution Eq.~(\ref{Eq.probability2}) is given by 
\be \label{CRB}
\vect{C}_{\rm CRB} = \frac{\vect{F}^{-1}}{m} = \frac{1}{m (F_{11} F_{22}-F_{12}^2)} 
\begin{pmatrix}
    F_{22} & -F_{12} \\
    -F_{12} & F_{11} \\
\end{pmatrix},
\ee
where $\vect{F}$ is the $2\times 2$ Fisher information matrix (FIM) with elements
\beq \label{Fisher}
\vect{F}_{k,k'} &=& \int_{-\pi/2}^{\pi/2} ds_1 \int_{-\pi/2}^{\pi/2} ds_2 \frac{1}{P_s\big(s_1, s_2 \vert \Xit \big)} \times \nonumber \\
&& \times \bigg( \frac{\partial P_s\big(s_1, s_2 \vert \vect{\Xi} \big)}{\partial \Xi_{k}} \bigg\vert_{\vect{\Xi}=\Xit}\bigg)  \bigg( \frac{\partial P_s\big(s_1, s_2 \vert \vect{\Xi} \big)}{\partial \Xi_{k'}} \bigg\vert_{\vect{\Xi}=\Xit} \bigg). \,\,\,\,\,\,
\eeq
In our case, $\vect{F}$ can be calculated numerically by using Eq.~(\ref{Eq.probability2}).
Analytical results can be obtained when the overlap between the different branches in Eq.~(\ref{Eq.probability2}) is relatively small and can be neglected. 
This happens in the limits $\sigmat \ll \pi$ and for $\sigmat \ll \thetat \ll \pi-\sigmat$.
Under these conditions, we obtain a diagonal FIM, 
\be \label{FIM}
\vect{F} = 
\begin{pmatrix}
    1/\sigmat^2 & 0 \\
    0 & 2/\sigmat^2 \\
\end{pmatrix}.
\ee
According to Eq.~(\ref{CRB}), for $m\gg 1$, the diagonal elements of $\vect{F}^{-1}$ provide
\be \label{Deltatheta} 
\Delta \thetae = \frac{\sigmat}{\sqrt{m}},
\ee
and
\be \label{Deltasigma}
\Delta \sigmae = \frac{\sigmat}{\sqrt{2m}}.
\ee
Equations (\ref{Deltatheta}) and~(\ref{Deltasigma}) coincides with Eq.~(\ref{results}), presented in the introduction with a lighter notation.
Equation~(\ref{Deltatheta}) has an intuitive justification.
We expect that the parameter $\thetat$ can be estimated efficiently from the position of one of the edges $A$-$D$ of the rectangular probability density, see Eq.~(\ref{Eq.coo}).
The width of the probability distribution at the edges is determined by the width of the uncorrelated noise when $\sigmat$ is sufficiently small~\cite{footnote1}.
We thus conclude that, for a narrow uncorrelated noise, the position of the vertices (and thus $\thetat$) can be determined with an uncertainty proportional to $\sigmat$, consistent with Eq.~(\ref{Deltatheta}).

Finally, by following the relation 
\be \label{sigma_uncorr2}
\sigmat^2 = \frac{1}{N_1} + \frac{1}{N_2 } + (N_1 + N_2) \taut^2, 
\ee
where $N_j$ is the (known) number of particles in the $j$th interferometer, we can infer the interaction parameter (assumed to be the same in both interferometers) as
\be \label{tauest}
\tau_{\rm est} = \sqrt{\frac{\sigmae^2}{N_1+N_2} -\frac{1}{N_1N_2}}.
\ee 
In this case, error propagation and Eq.~(\ref{Deltasigma}) provide 
\be \label{deltatauest}
\Delta \tau_{\rm est} = \frac{1}{N_1+N_2} \frac{\sigmat^2}{\taut} \frac{1}{\sqrt{2m} } \approx \frac{\taut}{\sqrt{2m}},
\ee
where the right-hand side approximation holds for $\taut \gg 1/(N_1 N_2)$.
Alternatively, an estimate of $\taut$ can be obtained by inverting 
\begin{equation} \label{sigma_uncorr3}
\sigmat^2 = \frac{4}{N_1^2} (\Delta \hat{J}_{y,1})^2 + \frac{4}{N^2_2} (\Delta \hat{J}_{y,2})^2,      
\end{equation}
which can be performed numerically.

\section{Joint estimation of differential phase shift and interaction strength}
\label{Sec4}

For a given value of $N_1$, $N_1$, $\taut$ (or equivalently $\sigmat$) and $\thetat$, we generate sequences of $m$ data, $\vect{z}_1$ and $\vect{z}_2$ according to Eq.~(\ref{Pz1z2quantum}).
For each $\vect{z}_1$ and $\vect{z}_2$ sequence, we estimate $\thetae$ and $\taue$, jointly, according to Eq.~(\ref{likelihood}) [we recall that $s^{(i)}_j = {\rm arcsin}(z_j^{(i)})$, with $i=1, ...,m$ and $j=1,2$]. 
%
We emphasize that the probability distribution ``generating'' the data, Eq.~(\ref{Pz1z2quantum}), is different from the one used to perform the ML estimation, Eq.~(\ref{Eq.probability2})~\cite{notaF}.
Nevertheless, as shown in Fig.~\ref{Figure1}, the two distributions, Eqs.~(\ref{Pz1z2quantum}) and (\ref{Eq.probability2}), match very well: this justifies the agreement between the numerical results reported below and Eqs.~(\ref{Deltatheta}) and (\ref{Deltasigma}).

\begin{figure}[t!]
\includegraphics[width=1\columnwidth]{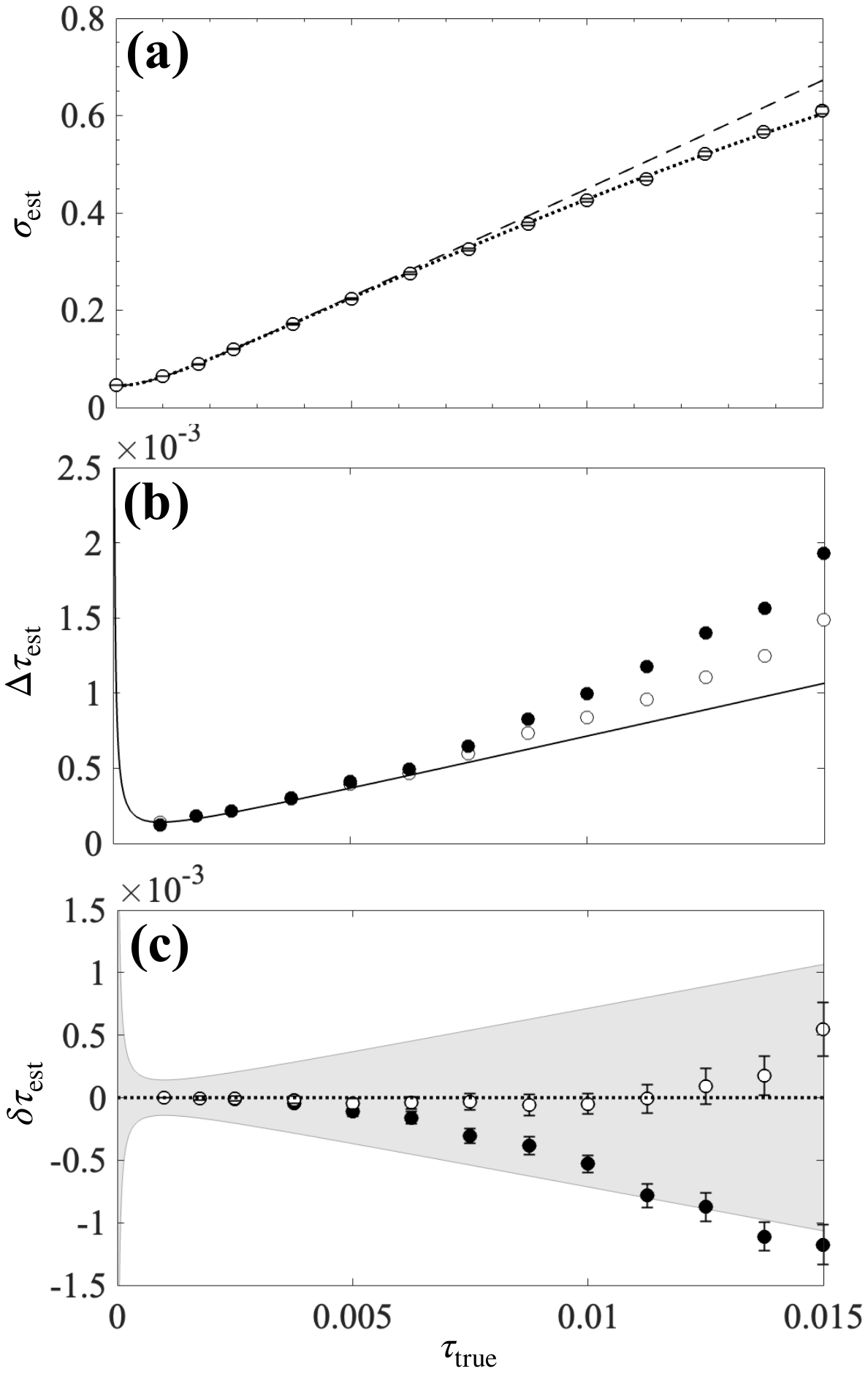}
\caption{
Panel (a) shows $\sigmae$ as a function of $\tau_{\rm true}$ (symbols).
The dashed line is Eq.~(\ref{sigma_uncorr2}), while the dotted line is Eq.~(\ref{sigma_uncorr3}). 
Panel (b) and (c) show $\Delta \taue$ and $\delta \taue$, respectively, as a function of $\taut$.
Dots refer the estimator Eq.~(\ref{tauest}), obtained through the numerical inversion of Eq.~(\ref{sigma_uncorr2}), while circles refer to the estimator obtained through the inversion Eq.~(\ref{sigma_uncorr3}).
The solid line in panel (b) is Eq.~(\ref{deltatauest}). 
The gray region in panel (c) is $\pm \Delta \taue$ around $\delta \taue = 0$ (dotted line), where $\Delta \taue$ is given by Eq.~(\ref{deltatauest}).
In all panels $N_1=N_2 = 1000$, $\thetat = 0.5\pi$ and $m=100$.
}
\label{Figure7}
\end{figure}

\begin{figure*}[t!]
\includegraphics[width=1\textwidth]{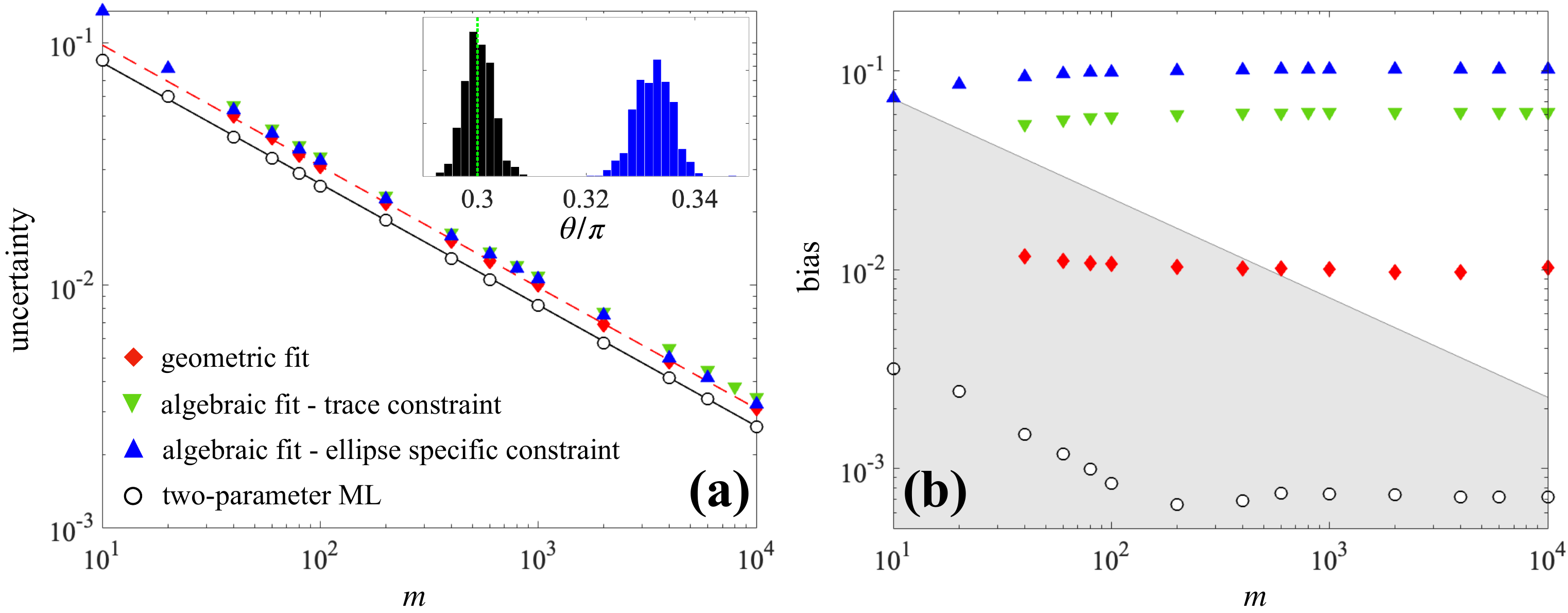}
\caption{Estimation uncertainty (a) and bias (b) as a function of $m$: circles refer to $\thetae$ obtained from the two-parameter ML approach, while the other symbols correspond to $\theta_{\rm elfit}$, obtained with different ellipse fitting methods (see footnote~\cite{notaelfit}).  
The CRB is shown in panel (a) by the solid line, while in panel (b) it delimits the gray region. 
In all panels, $N_1=N_2 = 10^3$, $\thetat/\pi = 0.3$ and $\taut = 0.005$.
Symbols are obtained by averaging over $10^3$ realizations. 
In some cases (algebraic fit with trace constraint and geometric fit) the fitting routine did not converge for relatively small values of $m$: this explains the lack of corresponding results for relatively small values of $m$.
The inset shows the statistical distribution of $\thetae$ (black histogram) and that of $\theta_{\rm elfit}$ (blue histogram, corresponding to ellipse specific constraint), compared to $\thetat$ (vertical dashed line).
Here $m=1000$.
}
\label{Figure8}
\end{figure*}

In Fig.~\ref{Figure6}(a) we show $\Delta \thetae$ as a function of the total number of particles in the two interferometers, $N_1+N_1$: different symbols refer to different values of $\taut$ (see legend).
The numerical results follow the CRB evaluated numerically (dashed line), which is well approximated by Eq.~(\ref{Deltatheta}) (solid line), with $\sigmat$ given by Eq.~(\ref{sigma_uncorr2}). 
In particular, for $\taut =0$ (black symbols and lines), the uncertainty of the ML estimate follows the standard quantum limit $\Delta \theta_{\rm est} = \tfrac{1}{\sqrt{m}}\big(\tfrac{1}{\sqrt{N_1}}+\tfrac{1}{\sqrt{N_2}}\big)$, decreasing with both the number of particles and the number of measurements.
Instead, for $\taut \neq 0$, the uncertainty bends up when increasing $N_1+N_2$, as a consequence of the second term in Eq.~(\ref{sigma_uncorr2}).
In Fig.~\ref{Figure6}(b) we show $\Delta \thetae$ as a function of $\thetat$, for $\taut=0$ (black dots) and $\taut=0.005$ (green triangles).
Consistently with panel~(a), the dashed line is the CRB, while the solid line is Eq.~(\ref{Deltatheta}).
The uncertainty is essentially independent from $\thetat$, except in a region of approximate width $\sigmat$, close to $\thetat=0$ (and $\pi$), where it becomes difficult to distinguish a small phase shift due to the finite width of the probability distribution.
This effect is also captured by the increase of the CRB.
In panels (c) and (d), we plot the bias $\delta \thetae = \bar{\theta}_{\rm est}-\thetat$ (symbols) as a function of $\thetat$, for $\taut=0$ and $\taut=0.005$, respectively.
In both panels, the gray region corresponds to a width given by the CRB, above and below the unbiased case $\delta \thetae=0$ (dashed line).
As we see, the bias is negligible compared to the estimator uncertainty, $\delta \thetae\ll \Delta \thetae$, over a broad range of $\thetat$ values. 
A significant bias rises only close to $\thetat=0,\pi$. 

In Fig.~\ref{Figure7}, we study the estimation of $\taut$.
In panel (a) we report $\sigmae$ as a function of $\tau_{\rm true}$ (circles).
The dashed line is Eq.~(\ref{sigma_uncorr2}), while the dotted line is Eq.~(\ref{sigma_uncorr3}): both equations agree for small values of $\tau$, as discussed above, while the latter reproduces well the numerical findings also for relatively large values of $\taut$.
In panel (b) and (c) we plot the uncertainty $\Delta \taue$ and the bias $\delta \taue=\bar{\tau}_{\rm est} - \taut$, respectively, as a function of $\taut$,
In both panels, dots correspond to the estimator Eq.~(\ref{tauest}), obtained through the inversion of Eq.~(\ref{sigma_uncorr2}), while circles corresponds to the estimator obtained by inverting Eq.~(\ref{sigma_uncorr3}).
The solid line in panel (b) is Eq.~(\ref{deltatauest}). 
%
%
Overall, the estimation of $\taut$ is optimal in an intermediate regime.  
On the one side, when $\taut$ is large (and the corresponding $\sigmat$ is large as well), the probability distribution becomes thick and it becomes increasingly difficult to estimate $\taut$.  
On the other side, if $\taut$ is small, namely $\taut \lesssim (N_1+N_2)/(N_1N_2)$, $\sigmat$ is dominated by the projection noise contribution and, also in this case, the uncertainty in the estimation of $\taut$ increases.

\subsection{Comparison with ellipse fitting}

As mentoned above, ellipse fitting techniques are usually considered for the analysis of differential interferometers. 
These methods consist of extracting the conic parameters $\vect{\nu}=\{a,b,c,d,e,f\}^\top$ of the curve 
\begin{equation}
\label{ellipsefit}
    a\, z_1^2+b\, z_1 z_2+c\, z_2^2+d\, z_1+e\, z_2+f=0
\end{equation}
that better fits the data.
Once the coefficients $\vect{\nu}$ are determined (see~\cite{CorgierARXIV} for a recent overview and footnote~\cite{notaelfit} for details about different fitting methods), $\theta$ is estimates as 
\begin{equation} \label{theta_ellipse_parameters}
    \theta_{\rm elfit}=\arccos\left(\frac{-b}{2\sqrt{ac}}\right),
\end{equation}
%
%
%
%
With this method, $\thetat$ is estimated by attempting to capture the behavior of the average $\bar{z}_1$ and $\bar{z}_2$, Eq.~(\ref{ellipse}), from scattered data, by using a fitting curve. 
In contrast, our ML approach includes -- by construction -- the information that the data fluctuate around the average.
%
We estimate, from the same set of measurement data $\vect{z}_1$ and $\vect{z}_2$, both $\thetat$ and $\sigmat$ (or equivalently $\taut$), while ellipse fitting only provides $\thetae$.
Similarly to ellipse fitting, our method does not require the full knowledge of the probability distribution Eq.~(\ref{Pz1z2quantum}).

To compare the ML approach with ellipse fitting, we extract both $\thetae$ and $\theta_{\rm elfit}$ from the same data $\vect{z}_1$ and $\vect{z}_1$ sampled according to Eq.~(\ref{Pz1z2quantum}).
For fixed $m$, the estimation is repeated approximately $10^3$ times to compute statistical uncertainties and the biases. 
%
%
The results are reported in Fig.~\ref{Figure8}.
In panel (a), we show the estimation uncertainty as a function of the number of measurements $m$.
%
%
%
%
%
%
In contrast to the ML approach (black circles), the uncertainty of ellipse fitting does not saturate the QCR. 
For instance, the results of a geometric fit (red diamonds) are approximately a factor 1.2 above the CRB. 
This factor varies with $\taut$, changing the width of the data distribution: for instance, in the case $\taut=0$, the uncertainty obtained with a geometric fit is a factor 1.4 above the CRB. 
In Fig.~\ref{Figure8}(b) we plot the estimation bias. 
The ML approach, is characterized by a bias $\delta \thetae$ that is approximately one orders of magnitude below that of the geometric fit and two orders of magnitudes below the uncertainty $\Delta \thetae$, in the range of measurements $100 \lesssim m \lesssim 1000$ that is relevant in current experiments. 
The small residual bias is due to the difference between the exact model Eq.~(\ref{Pz1z2quantum}) and the Gaussian approximation Eq.~(\ref{Eq.probability2}).
We can estimate that $\delta \thetae$ becomes comparable to $\Delta \thetae$ only for $m \gtrsim 10^5$ measurements.
The situation is well summarized in the inset of panel Fig.~\ref{Figure7}(a) where we plot the statistical distribution of $\thetae$ (black histogram) and that of $\theta_{\rm elfit}$ (blue histogram, corresponding to ellipse specific constraint), compared with the true value of the parameter, $\thetat/\pi = 0.3$ (vertical dashed line).
The width of the ML histogram is slightly smaller than that of the $\theta_{\rm elfit}$, but, most notably, the latter is centered away from the true value $\thetat$.
Similar results are observed for different values of $\thetat$ and $\taut$.


%
%
%

%
%
%
%

%
   
%

%
%

\section{Discussion and Conclusions}

In this manuscript, we have proposed a multiparameter ML approach for the joint  estimation of the differential phase shift as well as the width of the uncorrelated phase noise in a gradiometer configuration with coupled interferometers.
The analysis relies on a Gaussian model of uncorrelated noise that is well justified in the case of coupled Mach-Zehnder interferometers with interacting particles.
In this case, our methods allows to extract both the differential phase shift and the strength of the particle-particle interaction. 
Preliminary results show that our method can be also applied to estimate the differential phase shift with  sensitivities overcoming the standard quantum limit when using squeezed states in the differential interferometer. 
We leave a detailed analysis of this case~\cite{EcknerNATURE2023, CorgierARXIV} to future investigations. 
   
Several generalizations of our approach are possible.
First, it is interesting to consider other realistic sources of noise besides dephasing, such as fluctuations of the visibility and/or of the offset of the interference fringes, which are are main source of noise in several experiments.
Another relevant extension is the case of three or more interferometers working in parallel, which can be used to estimate gradients or spatially-varying signals~\cite{ZhengNATURE2022}.
It is also possible to extend the method to the case of correlated noise of final width.
In this case, it would be possible to estimate, simultaneously, both the correlated and uncorrelated noise width as well as the signal phase shifts in both interferometers (in the present manuscript the correlated noise has a flat distribution in the full $[-\pi, \pi]$ interval allowing the extraction of the differential phase only).

Finally, the approach outlined in this manuscript has been recently applied to estimate the differential phase shift and the uncorrelated noise width in a trapped gradiometer with atomic Bose-Einstein condensates~\cite{Petrucciani}.
In this experiment, the estimation of the interaction-induced dephasing has proved important to steer the gradiometer toward its optimal working point by tuning the contact interaction.

\acknowledgments
We thank R. Corgier, M. Malitesta, M. Prevedelli, L. Salvi, G. Rosi and G. Tino for discussions.
We acknowledge financial support by the project SQUEIS of the QuantERA ERA-NET Cofund in Quantum Technologies (Grant Agreement No. 731473 and 101017733) implemented within the European Unions Horizon 2020 Program. We also thank the financial support of the Italian Ministry of Universities and Research under the PRIN2022 project "Quantum sensing and precision measurements with nonclassical states". Finally the project has been co-funded by the European Union - Next Generation EU under the PNRR MUR project PE0000023-NQSTI and under the I-PHOQS 'Integrated Infrastructure Initiative in Photonic and Quantum Sciences'.


\begin{thebibliography}{100}
\bibitem{notaBS}
Interactions during the beam splitting operations can be neglected when the time required to implement the beam splitter is $t_{\rm BS} \ll \pi/(2N)$, see~\cite{PezzePRA2006}.
\bibitem{footnoteBS}
Without loss of generality, we model the two beam splitters of the interferometer as rotations around different axes. 
This guarantees that the mean interference signal has a sinusoidal dependence on the phase shift as in Eq.~(\ref{zave}).
\bibitem{notaP}
Equation (\ref{Eq.probability2}) is derived by first writing  
\begin{align} \label{Eq.probability}
P_s\big(s_1, s_2 \vert \theta, \sigma \big) &= \int_{-\pi}^{\pi} \frac{d \varepsilon}{2\pi} \int_{-\pi}^{\pi} d\chi~\mathcal{P}(\chi)~
\Big[\delta\big(\varepsilon - s_2 \big)+ \delta\big(\varepsilon  - \pi + s_2 \big) \Big] \nonumber  \\
& \,\, \Big[ \delta\big(\varepsilon + \chi + \theta - s_1\big) + \delta\big(\varepsilon + \chi + \theta - \pi + s_1\big) \Big],
\end{align}
and then computing the Kroenecker delta functions.
%
Equation~(\ref{Eq.probability2}) is normalized to 1, with each term of the sum equally contributing as $1/4$ to the integral $\int_{-\pi/2}^{\pi/2} ds_1 \int_{-\pi/2}^{\pi/2} ds_2 P_s\big(s_1, s_2 \vert \theta, \sigma \big) = 1$, for all values of $\theta$ and $\sigma$.
%
For $N\gg 1$, $s_1$ and $s_2$ can be considered as continuous variables.
\bibitem{footnote1}
For instance, considering the point $A$, we have 
\beq \label{PA}
&& P\big(s_1=-\pi/2, s_2=-\pi/2+\theta+\delta s_2 \vert \sigma_{\rm corr}, \sigma_{\rm uncorr}, \theta\big) = \nonumber \\
&& \qquad \qquad \qquad   = \frac{ P_{\rm uncorr}(2\theta +\delta s_2) + P_{\rm uncorr}(-\delta s_2) }{\pi}, 
\eeq
where $\delta s_2$ is a shift from the coordinate of Eq.~(\ref{Eq.cooA}).
If $\sigma_{\rm uncorr}$ is sufficiently small and $\theta$ large, the first term in Eq.~(\ref{PA}) can be neglected.
\bibitem{notaF}
For data $\vect{s}_1$ and $\vect{s}_2$ generated according to the probability $P_s(\vect{s}_1, \vect{s}_2 \vert \Xit) = \prod_{j=1}^m P_s(s_1^{(j)}, s_2^{(j)} \vert \Xit)$, the statistical properties of the ML guarantee that, for $m\gg 1$, $\bar{\vect{\Xi}}_{\rm est}$ converges to the true value $\Xit$ (namely $\delta\Xie \to 0$), and $\vect{C}$ equals the Cram\'er-Rao bound (CRB) for unbiased estimators~\cite{KayBOOK}.
%
In the analysis reported in Sec.~\ref{Sec4}, data $\vect{s}_1$ and $\vect{s}_2$ are generated according to Eq.~(\ref{Pz1z2quantum}) and thus the properties mentioned above are not guaranteed.
\end{thebibliography}
\end{document}